\documentclass[useAMS,usenatbib,usegraphicx,twocolumn]{mn2e}
  
\usepackage{float}
\usepackage{aas_macros}
\usepackage{threeparttable}
\usepackage{grffile}
\usepackage{amsmath, amssymb}
\usepackage{multirow}
\usepackage{wallpaper}
\usepackage{mathbbol}
\usepackage{rotating}
\usepackage{blindtext}
\usepackage[T1]{fontenc}
\usepackage{changes}
\usepackage{lscape}

\usepackage{tocloft}
\newcommand{\juan}[1]{{\bf \color{magenta} #1}}
\renewcommand{\juan}[1]{#1}

\newcommand{\vek}{\underline}
\newcommand{\fat}{\textbf}
\newcommand{\ita}{\textit}
\newcommand{\del}{\partial}

\newcommand{\beq}{\begin{equation}}
\newcommand{\eeq}{\end{equation}}  
\newcommand{\RNum}[1]{\uppercase\expandafter{\romannumeral #1\relax}}

\newcommand{\bk}[1]{{\bf \textcolor{blue}{#1}}}
\renewcommand{\bk}[1]{#1}
  \title[HRO in turbulent media]{The relative orientation between the magnetic field and gas density structures in non-gravitating turbulent media}
  
  \author[B.~K\"ortgen \& J.~D.~Soler]{Bastian~K\"ortgen$^1$ and Juan~D.~Soler$^2$\\
  $^{1}$ Hamburger Sternwarte, Universit\"at Hamburg, Gojenbergsweg 112, D-21029 Hamburg, Germany \\
  $^{2}$ Max Planck Institute for Astronomy, K\"onigstuhl 17, D-69117 Heidelberg, Germany
  }

\date{Released 2020}

\pagerange{\pageref{firstpage}--\pageref{lastpage}} \pubyear{2020}

\begin{document}

\label{firstpage}
\maketitle

\begin{abstract}
Magnetic fields are a dynamically important agent for regulating structure formation in the interstellar medium.
The study of the relative orientation between the local magnetic field and gas (column-) density gradient has become a powerful tool to analyse the magnetic field's impact on the dense gas formation in the Galaxy.
In this study, we perform numerical simulations of a non-gravitating, isothermal gas, where the turbulence is driven either solenoidally or compressively. 
We find that only simulations with an initially strong magnetic field (plasma-$\beta<1$) show a change in the preferential \juan{orientation} between the magnetic field and isodensity contours, from mostly parallel at low densities to mostly perpendicular at higher densities.
Hence, compressive turbulence alone is not capable of inducing the transition observed towards nearby molecular clouds.
At the same high initial magnetisation, we find that solenoidal modes produce a sharper transition in the relative orientation with increasing density than compressive modes.
We further study the time evolution of the relative orientation and find that it remains unchanged by the turbulent forcing after one dynamical timescale. 
\end{abstract}
\begin{keywords}
galaxies: magnetic fields; galaxies: ISM; ISM: magnetic fields; ISM: clouds; stars: formation
\end{keywords}

\section{Introduction}\label{sec:intro}

Magnetic fields are an essential component of the interstellar medium \citep[ISM,][]{ferriere2001,crutcher2012,klessenANDglover2016}.
They play an important role in shaping the interstellar gas in and around star-forming clouds, by favouring the formation of filamentary structures and reducing the number of overdensities that can form new stars 
\citep[see][for a recent review]{hennebelleANDinutsuka2019,pattleANDfissel2019}.
However, its exact influence on the cycle of matter in the ISM remains to be better understood.

Observations indicate that the magnetic energy density is in rough equipartition with the other energy densities in the local ISM \citep{heilesANDcrutcher2005,Beck15}.
Gas motions are unaffected when they propagate parallel to the mean direction of the field ($B_{0}$), but strongly inhibited by the magnetic pressure when they propagate normal to $B_{0}$ \citep{field1965}.
Consequently, the accumulation of gas that precedes the formation of molecular clouds (MCs) is most effective only if the perturbations propagate almost parallel to $B_{0}$ \citep{hennebelle2000,hartmann2001,Koertgen15,Koertgen18L}. 
This implies that magnetic fields are a source of asymmetry in the accumulation of interstellar gas from the diffuse ISM. 

Such an asymmetry in the distribution of the over-densities in the ISM has been observed toward MCs in the relative orientation between the gas column densities and the plane-of-the-sky magnetic field orientation, inferred from the dust polarised emission observations  \citep{planck2015-XXXV,fissel2019,soler2019}.
Using assumptions on the mapping of the gravitational potential by the distribution of submillimeter emission, \citet{Koch12} use this asymmetry to infer the local magnetic field strength and its relative importance to other forces in the analysis of SMA polarisation data.

The study of numerical simulations of magnetohydrodynamic (MHD) turbulence in MCs reveals that the relative orientation between density structures ($\rho$) and the magnetic field $\vec{B}$ is related to the magnetisation of the gas \citep{Soler13}.
Isothermal simulations of turbulent MCs show that when the kinetic energy density is larger than the magnetic energy density, $\rho$ structures are aligned to $\vec{B}$ across all $\rho$ values. 
In contrast, if the magnetisation is high, the $\rho$ structures change orientation from mostly parallel to mostly perpendicular with increasing $\rho$.
This result has also been found in simulations of dense core formation in colliding flows \citep{Chen16} and, more recently, in MCs selected from 1-kpc-scale multiphase numerical simulations \citep{seifried2020}.

There are several explanations for the origin of this transition from parallel to perpendicular orientation with increasing density.
\citet{Chen16} argue that the transition happens once the flow becomes super-Alfv\'enic due to gravitational collapse. 
More recently, \citet{seifried2020} found that the change in relative orientation correlates well with the regime, where the gas becomes magnetically supercritical. 
These two findings are quite similar since the gas becomes gravitationally unstable and magnetically supercritical at roughly the same column density \citep{Vazquez11a}.

\cite{Soler17b} developed a theory to describe the evolution of the angle between the magnetic field and the gas structures. 
They showed that a perpendicular arrangement of magnetic fields and gas structures is a consequence of gravitational collapse or, more generally, converging gas motions.
Other interpretations, such as those presented in \cite{hu2019}, assign the anisotropies predicted by the theory of interstellar MHD turbulence as the source of the observed relative orientation between $\rho$ structures and $\vec{B}$ \citep{goldreich1995}.

In this work, we investigate the role turbulence plays in \bk{setting} the relative orientation between the density structures and the magnetic field, mainly focusing on the role of compressive and solenoidal modes.
For this purpose, we analyse a set of numerical simulations of driven MHD turbulence in an isothermal gas without an external \juan{gravitational} potential and self-gravity.
Using this numerical experiment, we aim to identify to what extent the observed trends in relative orientation between $\rho$ and $\vec{B}$ is explained by the influence of the different modes of turbulence in the ISM.

This paper is organised as follows.
Sec.~\ref{sec:method} briefly reviews the basic methodology behind the analysis presented below.
Sec.~\ref{sec:sims} describes the used numerical code as well as the initial conditions for this study. 
Finally, in Sec.~\ref{sec:results}, we present and discuss our findings. 
This study is closed with a summary of our main findings in Sec.~\ref{sec:summary}.

\section{Methodology}\label{sec:method}

One method for quantifying the orientation of the density structures is the characterisation through its gradient, which relies on the fact that the density gradients ($\nabla\rho$) are normal to the iso-density contours \citep{Soler13}. 
This method has the advantage of providing a direct connection to quantities in the MHD equations.
Using this connection, \cite{Soler17b} derived the time evolution of the orientation between the magnetic field and the gas density gradient (denoted by the angle $\phi$), which is 
\beq
\frac{\mathrm{dcos}\phi}{\mathrm{d}t}=C+\left(A_1+A_{23}\right)\mathrm{cos}\phi, 
\label{eq:hro}
\eeq
where the terms are
\beq
C=-\frac{\del_\mathrm{i}\left(\del_\mathrm{j}v_\mathrm{j}\right)}{\left(R_\mathrm{k}R_\mathrm{k}\right)^{1/2}}b_\mathrm{i},
\eeq
\beq
A_1=\frac{\del_\mathrm{i}\left(\del_\mathrm{j}v_\mathrm{j}\right)}{\left(R_\mathrm{k}R_\mathrm{k}\right)^{1/2}}r_\mathrm{i},
\eeq
\beq
A_{23}=\frac{1}{2}\left(\del_\mathrm{i}v_\mathrm{j}+\del_\mathrm{j}v_\mathrm{i}\right)\left[r_\mathrm{i}r_\mathrm{j} - b_\mathrm{i}b_\mathrm{j}\right].
\eeq
Here, $v_i$ denotes the fluid velocity,
\beq
r_i \equiv \frac{R_{i}}{\left(R_{k}R_{k}\right)^{1/2}},
\eeq
\bk{with} $R_{i}\equiv\del_\mathrm{i}\log\rho$, and
\beq
b_i \equiv \frac{B_{i}}{\left(B_{k}B_{k}\right)^{1/2}},
\eeq
which are the unit vectors of the gradient of the (logarithmic) density and the magnetic field, respectively. 
These terms highlight the crucial role of converging and diverging gas motions, represented by the diagonal elements of the shear tensor, $\del_\mathrm{i}v_\mathrm{j}$, and their link to the directions of the magnetic field and density gradients.
Thus, in general, any process that leads to a spatial change in the velocity field can induce a transition in the relative orientation between the magnetic field and gas density structures.
We note that, as discussed in \citet{seifried2020}, any coefficient in Eq.~\eqref{eq:hro} can be responsible for a change in relative orientation.
However, in standard scenarios that are not related to strong shocks, the terms related to $\del_\mathrm{i}\left(\del_\mathrm{j}v_\mathrm{j}\right)$ tend to be negligible.
That leaves the term $A_{23}$, which is related to the velocity strain tensor and the symmetric tensors, as the dominant term in Eq.~\eqref{eq:hro}.

\section{Numerical simulations}\label{sec:sims}
\subsection{Initial conditions}

We set a cubic box with edge length \mbox{$L=10\,\mathrm{pc}$}. 
The domain is filled with gas with initial number density \mbox{$n_\mathrm{init}=536\,\mathrm{cm}^{-3}$}, which gives a total mass of 
\mbox{$M_\mathrm{tot}\sim3\times10^4\,\mathrm{M}_\odot$}. 
The gas is isothermal at a temperature of \mbox{$T=11\,\mathrm{K}$}.
We use a mean molecular weight $\mu_\mathrm{mol}=2.4$, both typical for dense molecular gas. 
The sound speed associated with this temperature is \mbox{$c_\mathrm{s}=0.2\,\mathrm{km/s}$}. 
These conditions are similar to those studied in \cite{Soler13}.

Since we are interested in the relative orientation of the magnetic field with respect to the density gradient, we set up an initially homogeneous magnetic field $\fat{B}=B_0\hat{\fat{x}}$, where the field strength $B_0$ is calculated from the initial ratio of thermal to magnetic pressure 
\beq
\beta=\frac{P_\mathrm{th}}{P_\mathrm{b}}=\frac{nk_\mathrm{B}T}{B_0^2/8\pi},
\eeq
where $k_\mathrm{B}$ is Boltzmann's constant.
The turbulence in the domain is driven by an Ornstein-Uhlenbeck process with peak wavenumber $k=2$, which corresponds to half the box size, i.e. $L/2=5\,\mathrm{pc}$ \citep[see e.g.][]{Federrath08,Federrath09a,Federrath10b}. 
The contribution from solenoidal or compressive modes is determined by the parameter $\zeta_\mathrm{d}$, which enters the projection tensor
\beq
\mathcal{P}_\mathrm{ij}=\zeta_\mathrm{d}\left(\delta_\mathrm{ij}+\frac{k_\mathrm{i}k_\mathrm{j}}{\left|k\right|^2}\right)+\left(1-\zeta_\mathrm{d}\right)\frac{k_\mathrm{i}k_\mathrm{j}}{\left|k\right|^2}.
\eeq
We choose only the two extreme cases of \juan{purely} solenoidal driving with $\zeta_\mathrm{d}=1$ or \juan{purely} compressive driving with $\zeta_\mathrm{d}=0$. 
The amplitude of the turbulence forcing term is adjusted in such way that the sonic Mach number in the quasi-stationary state is \mbox{$\mathcal{M}_\mathrm{s}\sim7.5$}. 
An overview of the initial conditions is given in \mbox{Tab.~\ref{tab:ic}}.
\begin{table}
    \centering
    \caption{List of performed simulations. $\beta$ denotes the initial 
    ratio of thermal to magnetic pressure and the sonic and alfv\'enic 
    Mach numbers depict the ratio of kinetic to thermal and kinetic to 
    magnetic energy, respectively.}
    \begin{tabular}{lcccc}
        \hline
        \hline
         \fat{Run name} &$\beta$   & $\mathcal{M}_\mathrm{sonic}$   &$\mathcal{M}_\mathrm{alfvenic}$ &\fat{Driving}\\
         \hline
         B10com     &10 &7.5    &16.7   &compressive\\
         B10sol     &10 &7.5    &16.7   &solenoidal\\
         \hline
         B1com      &1  &7.5    &5.3    &compressive\\
         B1sol      &1  &7.5    &5.3    &solenoidal\\
         \hline
         B0.01com   &0.01   &7.5    &0.5    &compressive\\
         B0.01sol   &0.01   &7.5    &0.5    &solenoidal\\
         \hline
         \hline
    \end{tabular}
    \label{tab:ic}
\end{table}

\subsection{Numerics}
The simulations were performed using the Eulerian finite volume code \textsc{flash} \citep[version 4.6.1,][]{Dubey08}. 
During each timestep, \textsc{flash} solves the equations of ideal magnetohydrodynamics using a positive-definite 5-wave Riemann solver \citep{Bouchut09,Waagan11}. 
Periodic boundary conditions are applied at all domain walls. 
The root grid is at a resolution of $64^3$ and we allow for localised adaptive refinement up to an effective resolution of $512^3$. 
Despite the lack of self-gravity of the gas, the grid is adaptively refined once the local Jeans-length, $\lambda_\mathrm{J}$, is resolved by less than eight cells. 
In addition, de-refinement is applied to regions, where $\lambda_\mathrm{J}$ is resolved by more than 16 grid cells.
\section{Results}\label{sec:results}
We commence with analysing the relative orientation of the magnetic field and the 
density gradient. 
In what follows and not otherwise noted, all data are analysed at 1.5 turn-over times, i.e. in a state in which the turbulent fluctuations can be thought of as in steady-state.
\subsection{Overview of the \juan{simulation scenarios}}
\begin{figure*}
    \begin{tabular}{lr}
    \includegraphics[width=0.5\textwidth]{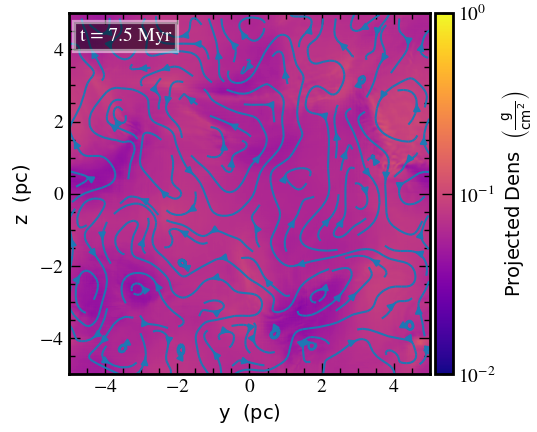}&\includegraphics[width=0.5\textwidth]{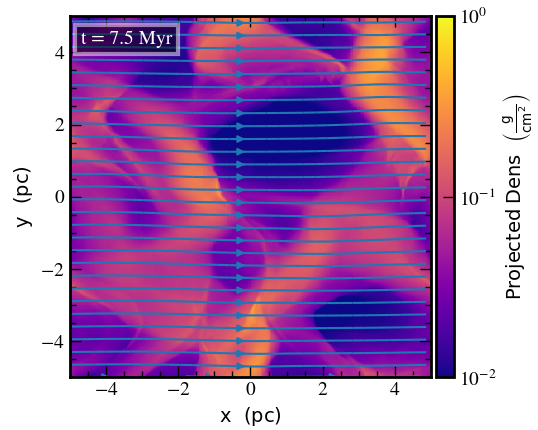}\\
    \includegraphics[width=0.5\textwidth]{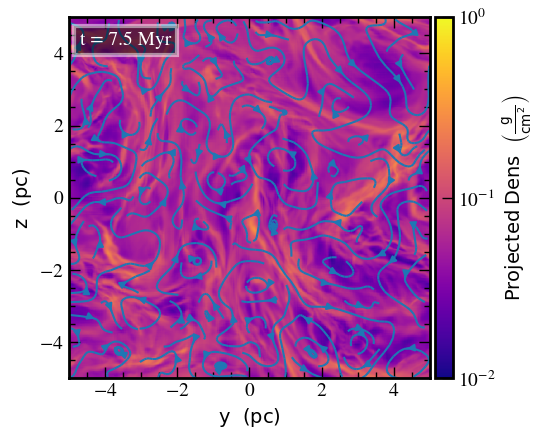}&\includegraphics[width=0.5\textwidth]{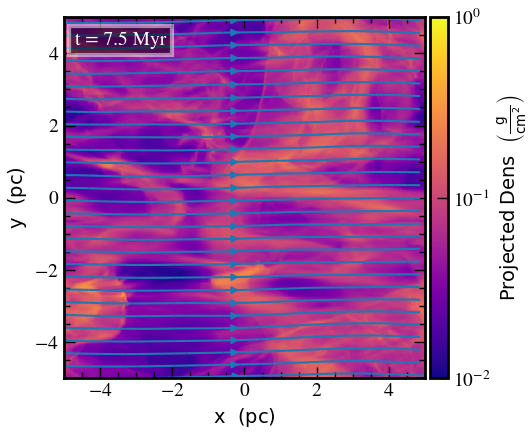}\\
    \includegraphics[width=0.5\textwidth]{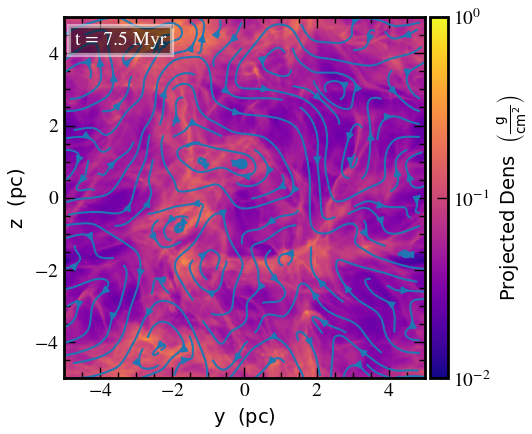}&\includegraphics[width=0.5\textwidth]{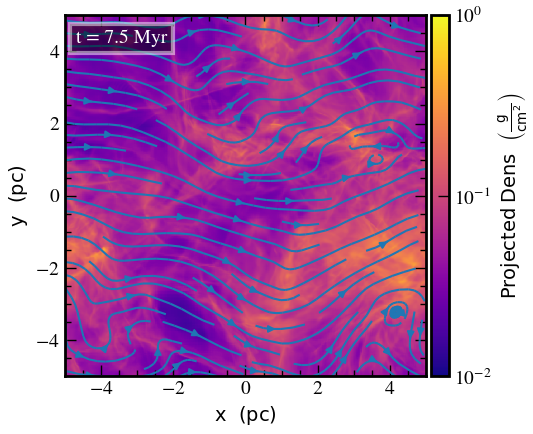}
    \end{tabular}
    \caption{Surface density maps at $t=1.5\,t_d$ along 
    the x-direction (i.e. parallel to the initial \bk{magnetic field direction}, left) and the z-direction (perpendicular to the initial field \bk{direction}, right) for \bk{the} runs B0.01com (top), B0.01sol (middle) and B1sol (bottom).}
    \label{fig:cdens}
\end{figure*}

\bk{We present in Fig.~\ref{fig:cdens} surface density maps for the runs {B0.01com}, {B0.01sol} and {B1sol}.} 
The projection along two different axes reveals pronounced differences. 
\juan{T}he density structure varies for different driving. 
The \bk{case of} compressive driving shows a few regions of enhanced density with the major axis primarily perpendicular to the background magnetic field. 
The density contrast vanishes almost completely after projection along the background magnetic field. 
\juan{We also note} the different morphology of the magnetic field. 
\juan{S}olenoidal turbulence driving results in many more substructure, which closely resemble striations\juan{, such as those described in \cite{tritsis2016,chen2017,beattie2020}}. 
\juan{These} striations are more pronounced in the case with small plasma-$\beta$, since here the magnetic field is strong enough to guide the flow of gas. 
This can also be inferred from the magnetic field morphology. 
While for the low-$\beta$ scenarios, the initial field direction is retained, it is largely disturbed in the $\beta=1$ case. 
However, the initial \juan{magnetic field} direction is still dominant.
\subsection{The relative orientation parameter}
\begin{figure*}
    \includegraphics[width=0.5\textwidth,angle=-90]{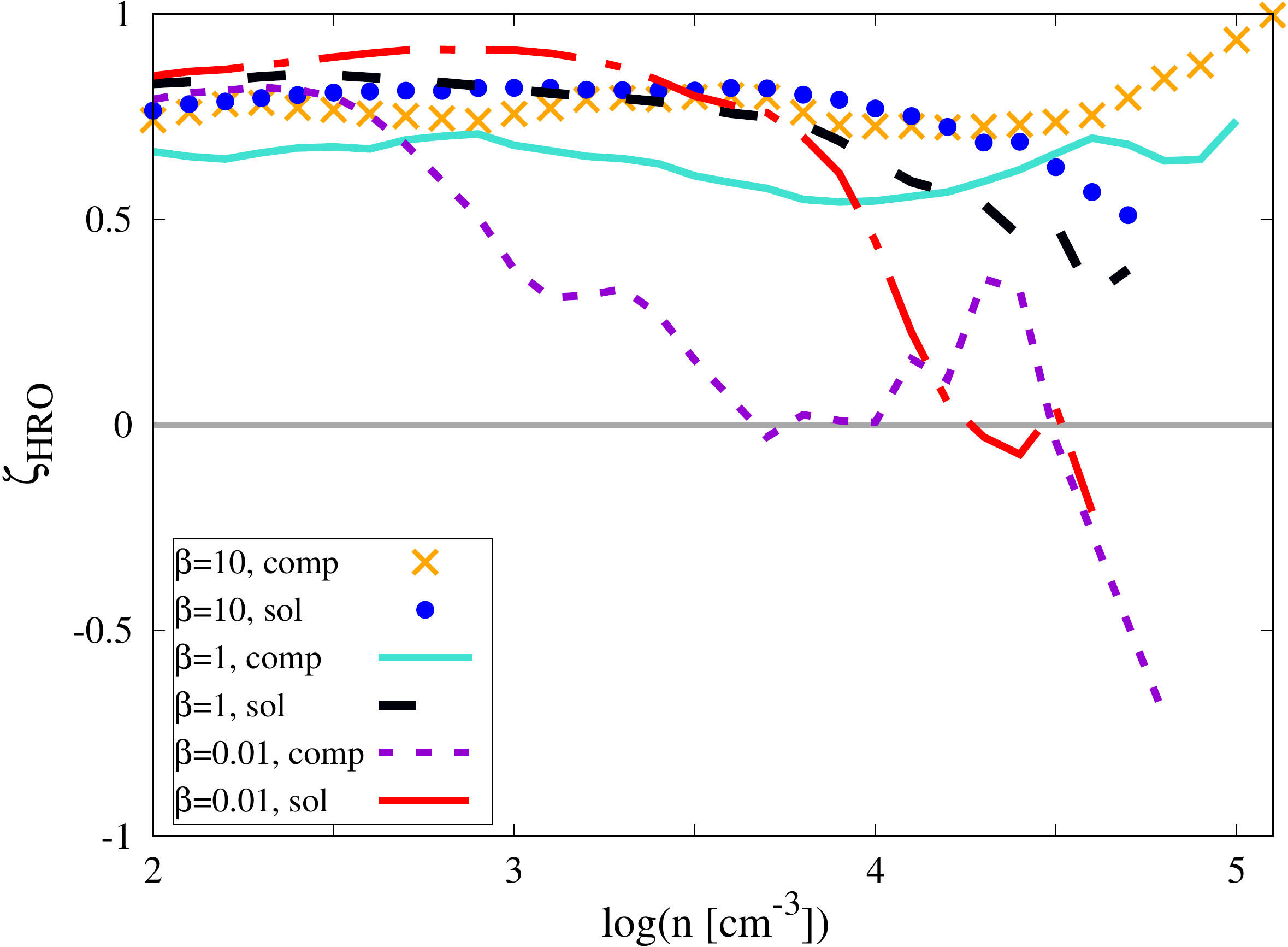}
    \caption{The HRO shape parameter as a function of number density for the different simulations. 
    While the weakly magnetised scenarios show no change in relative orientation, the highly magnetised 
    cases indicate a transition from preferentially perpendicular to parallel alignment \bk{(in terms of the density gradient)}. Note further 
    that the decrease in the shape parameter occurs at smaller densities for the compressive case, 
    although there appears some density regime of no preferred orientation.}
    \label{fig:hro}
\end{figure*}

\juan{We quantif\bk{y} the relative orientation by using the histogram of relative orientation (HRO) shape parameter, a quantity introduced in \cite{Soler13} and defined as
\beq
\zeta_\mathrm{HRO}\equiv\frac{A_{\perp}-A_{||}}{A_{\perp}\bk{+}A_{||}},
\eeq
where $A_{||}$ and $A_{\perp}$ correspond to the counts in the histogram \bk{of} $\cos\phi$ in the ranges that \bk{represent} parallel \bk{or} perpendicular \bk{alignment of} $\nabla\rho$ and $\vec{B}$.
Following the ranges introduced in \cite{Soler13} and \citet{seifried2020}, we have used $|\cos\phi|<0.25$ for \bk{$A_{\perp}$} and $|\cos\phi|>0.75$ for \bk{$A_{\parallel}$}.
The definition of the histogram in terms of $\cos\phi$ acknowledges the fact that, in 3D, the distribution of pairs of randomly oriented vectors is flat in terms of $\cos\phi$ and not in $\phi$.
Given the aforementioned definitions, $\zeta_\mathrm{HRO}$\,$>$\,$0$ if the density structures are mostly parallel to the magnetic field \bk{(or $\nabla\rho\perp\vek{B}$)} and $\zeta_\mathrm{HRO}$\,$<$\,$0$ if the density structures are mostly perpendicular to the magnetic field \bk{($\nabla\rho\parallel\vek{B}$)}.
A homogeneous distribution of relative orientation angles would correspond to $\zeta_\mathrm{HRO}$\,$\approx$\,$0$.
We note that more sophisticated methods based on circular statistics have been introduced to characterise the relative orientations in 2D, for example in \cite{jow2018} and \cite{soler2019}, but its implementation in 3D distributions of angles is still work in progress.
}\\
\juan{W}e show $\zeta_\mathrm{HRO}$ in multiple bins of gas number density $n$ in Fig.~\ref{fig:hro}. 
Two \juan{main} features are observed. 
\juan{First,} the \juan{values of $\zeta_\mathrm{HRO}$\,$\approx$\,$1$ across $n$} for the weakly magnetised scenarios with $\beta=\left\{1,10\right\}$. 
This is expected since the magnetic field is sub-dominant in these situations and dragged along with the flow of gas. 
\juan{Therefore,} it imposes no asymmetry to the fluid flow.
More importantly, there are only minor differences between solenoidally and 
compressively driven turbulence cases. 
This means that compressive driving of the turbulence alone, i.e. due to stellar feedback, is not enough to induce a change in relative orientation. 
In agreement with previous studies of magnetised (and self-gravitating) media \citep{Soler13,Soler17b,seifried2020}, a dynamically dominant magnetic field must be present.\\
\juan{Second,} the decrease of $\zeta_\mathrm{HRO}$ with density for the smallest values of $\beta$. 
This behaviour is typical in such scenarios, since here the magnetic field is strong enough to guide the gas flow. 
This leads to compression of gas along the field lines and a subsequent build-up of a density gradient parallel to the field.
Among the two cases, the one with compressive turbulence shows the earlier decrease of $\zeta_\mathrm{HRO}$. 
This is due to the naturally enhanced divergence of the velocity field, which is zero by definition in the solenoidal case. 
However, the magnetic and density gradient fields show a tendency for almost no preferred orientation \bk{($\zeta_\mathrm{HRO}\approx0$)} after it has started to decrease. 
There is even a slight re-increase observed at around $\mathrm{log}\left(n\right)\sim4.4$, before $\zeta_\mathrm{HRO}$ transitions towards negative values.
The fact that the solenoidally driven case shows a decreasing $\zeta_\mathrm{HRO}$ as well is due to the non-linear conversion of solenoidal to compressive modes in the velocity field \citep{Konstandin12}.
However, as this conversion is only efficient enough in the denser gas the start of the decrease naturally appears at higher densities.
\juan{This effect} can be associated with smaller scales that are not influenced by the driving on large scales.\\
In \juan{summary}, the strongest magnetisation yields the expected transition from preferentially perpendicular to parallel alignment. 
\juan{When considering the relative orientation at the highest densities ($n$\,$>$\,$10^{4}$\,cm$^{-3}$), the relative orientation is the same for the solenoidal and compressive forcing and seems to solely depend on the magnetisation.}
\juan{At lower densities ($10^{2}$\,$<$\,$n$\,$<$\,$10^{4}$\,cm$^{-3}$), there appears some difference between the solenoidal and compressive forcing}.
\subsection{What determines \bk{the change in relative orientation?}}
\begin{figure}
    \includegraphics[width=0.35\textwidth,angle=-90]{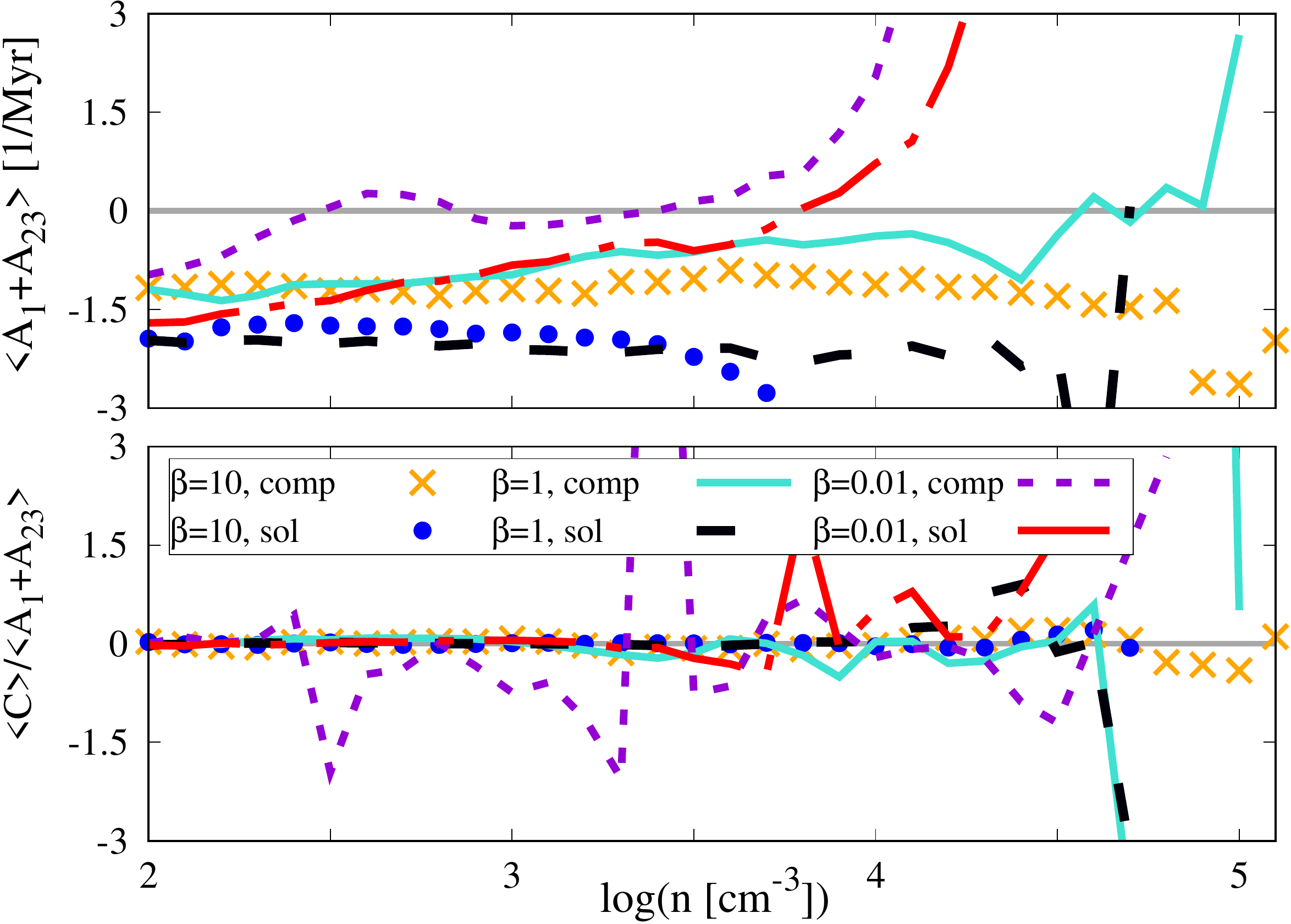}
    \caption{\bk{Density dependence of the coefficients entering Eq.~\eqref{eq:hro}.} \ita{Top:} Average of the sum of the A-terms. \ita{Bottom:} The ratio of the average C-term 
    and the total A-term.}
    \label{fig:terms_tot_ratio}
\end{figure}

The form of Eq.~\eqref{eq:hro}, provided in Sec.~\ref{sec:method}, indicates that, if $A_1+A_{23}\equiv A>C$, the evolution of the relative orientation depends on the sign of the $A$-coefficient and thus on the relative importance of the change in velocity divergence and the strain in the \bk{velocity field} \citep{Soler17b}.
The resulting density-dependence of $A$ and $C$ is shown in Fig.~\ref{fig:terms_tot_ratio}. At the lowest densities, the $A$-term is negative, which indicates a transition towards 
$\mathrm{cos}\,\phi\rightarrow0$ according to Eq.~\eqref{eq:hro}, as long as $C$ is small. Major deviations in the evolution, however, start to appear already from $\mathrm{log}\left(n\right)\sim2.5$ on. These deviations are 
furthermore not only dependent on the magnetisation of the gas, but also depend on the way the turbulence is 
driven. While \bk{the $A$-term in} the scenarios \ita{B10sol} and \ita{B1sol} become\bk{s} even more negative with density, 
the corresponding \bk{terms in the runs with} compressive turbulence stay almost constant, with a slight increase 
towards $A\sim0$ at the highest densities\footnote{The maximum densities around $\mathrm{log}\,n\sim5$ 
are characterised by low-number statistics and thus should be interpreted with caution.}. In contrast, 
the two \bk{strongly magnetised} cases with $\beta=0.01$ show a transition from $A<0$ to $A>0$. As predicted by 
\citet{Soler17b}, a transition to $\zeta_\mathrm{HRO}<0$ is achieved as soon as the $A$-term 
changes sign. The change in sign in our case is accompanied by a decrease of 
$\zeta_\mathrm{HRO}$, rather than a transition to below zero. The density regime where 
$\zeta_\mathrm{HRO}\sim0$ for scenario \ita{B0.01com}, \bk{which indicates no preferred orientation}, is described by $A\sim0$ and thus $C>A$ (see 
Fig.~\ref{fig:terms_tot_ratio}, bottom panel). \bk{The corresponding 
simulation with solenoidal forcing crosses the zero-point line at slightly higher densities.} Both strongly magnetised runs then show 
a sharply increasing $A$ as a function of density.\\
\bk{To sum up}, the density dependence of the $A$-term is controlled by the magnetisation of the gas in 
terms of $A$ transitioning from negative to positive values. On the other hand, for equipartition 
or very weak fields ($\beta=1$ or $\beta=10$), the evolution appears more likely to be controlled by 
the type of turbulent forcing. For solenoidal forcing, $A$ becomes more negative, while for 
compressive forcing it stays rather constant or shows a shallow positive slope as a 
function of density.
\subsection{Fluid strain or projected divergence?}
\begin{figure}
    \includegraphics[width=0.25\textwidth,angle=-90]{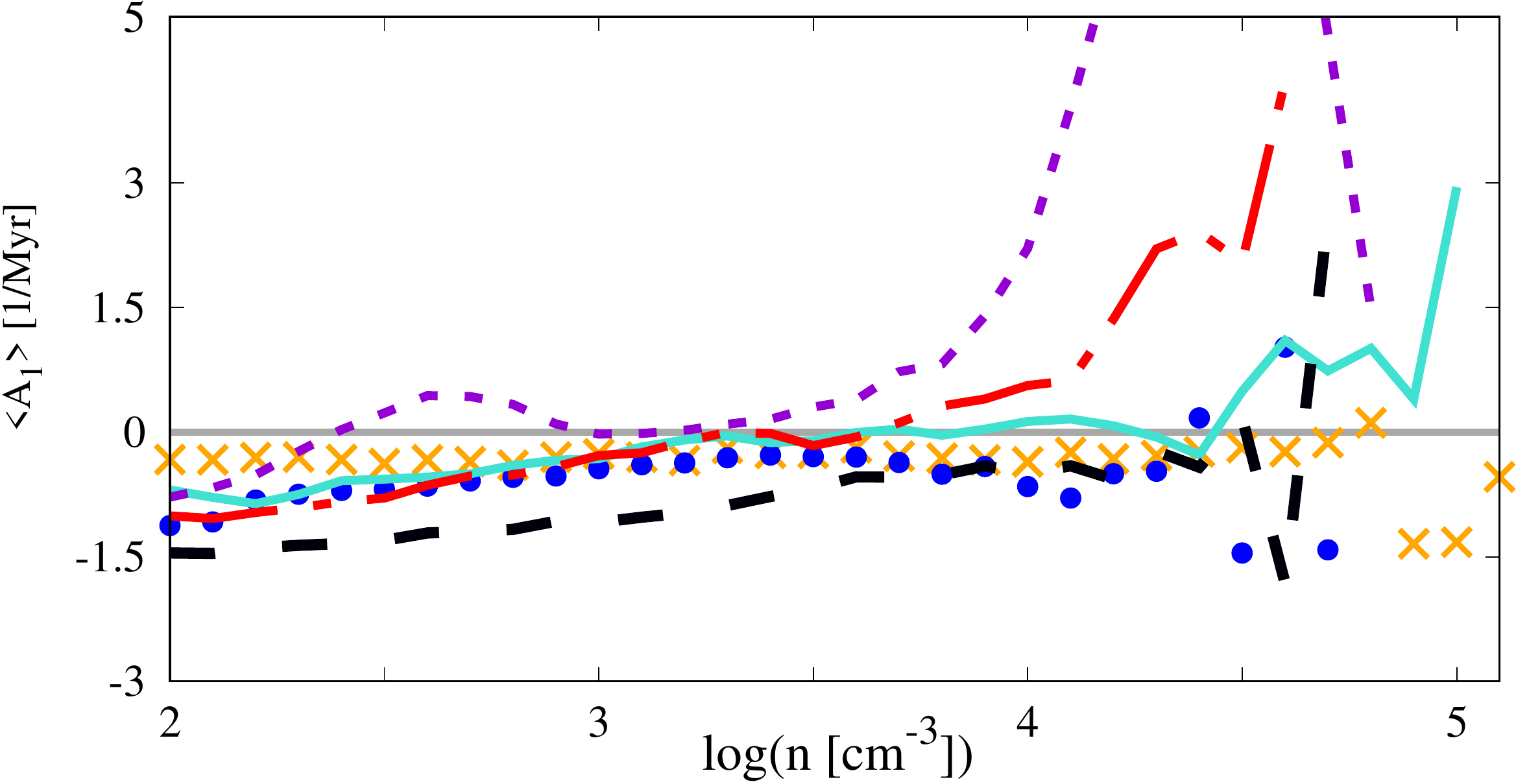}\\
    \includegraphics[width=0.25\textwidth,angle=-90]{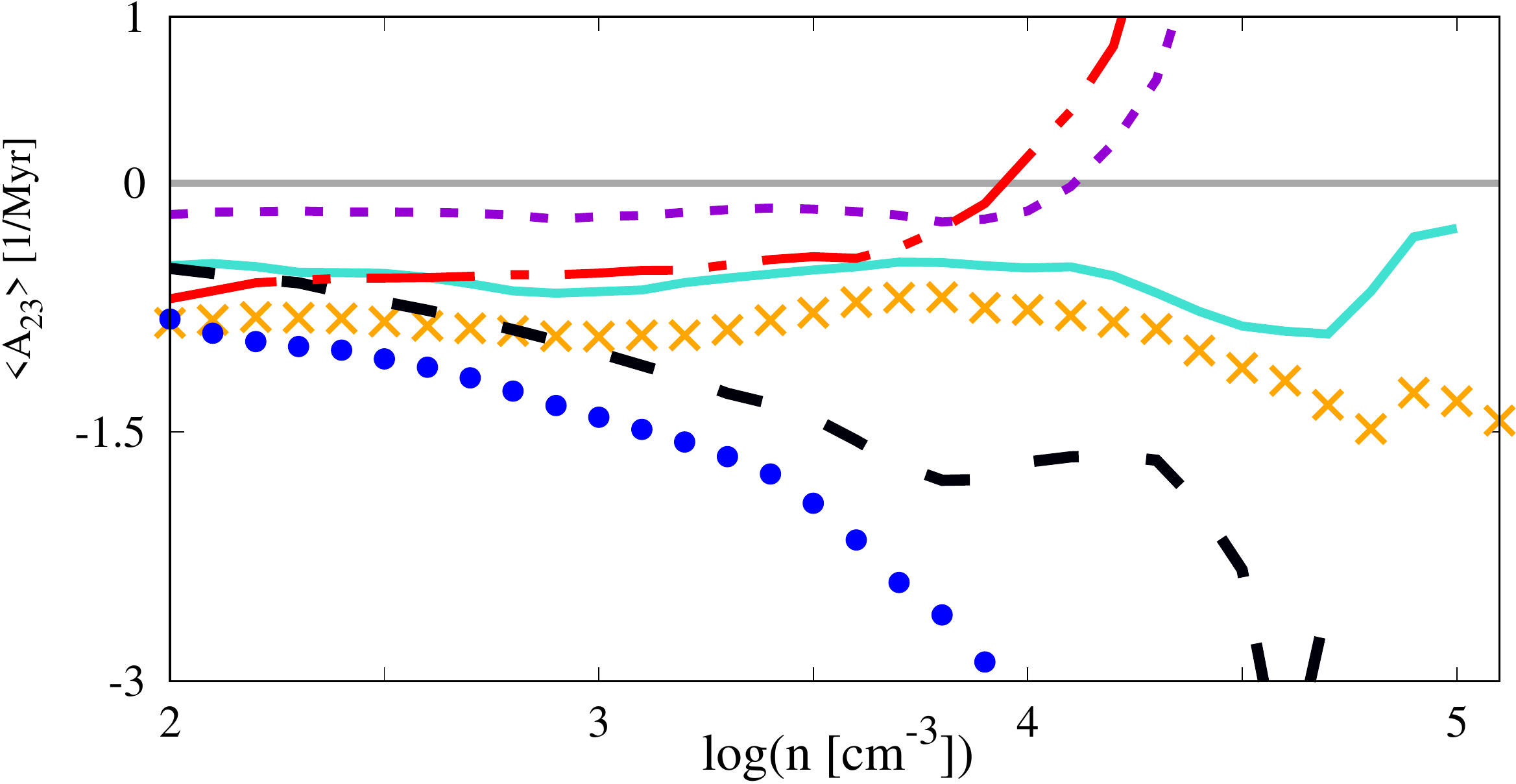}\\
    \includegraphics[width=0.25\textwidth,angle=-90]{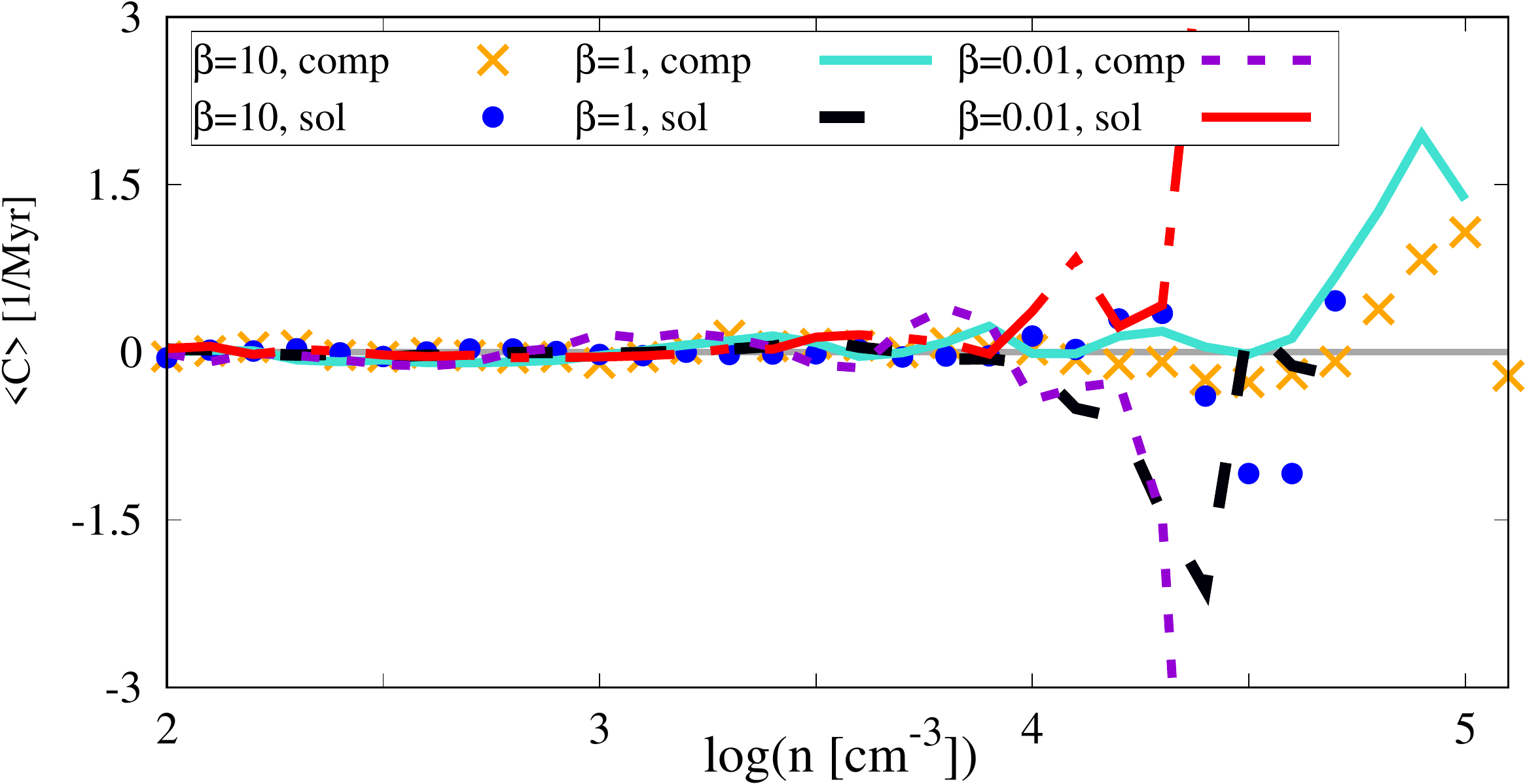}\\
    \caption{Density dependence of the various \bk{individual} coefficients, which 
    control the rate of change of the relative orientation. We 
    note that we only show the average values here and that the 
    uncertainty can be large \citep{seifried2020}. Please note the 
    different scaling of the y--axes.}
    \label{fig:terms_ind}
\end{figure}

Above it was shown that $A>C$ in most cases. Here, we discuss which part of $A$ is responsible for 
the transition from $A<0$ to $A>0$ and thus for a change in relative orientation from preferentially 
perpendicular alignment to almost parallel alignment\footnote{This statement 
applies to the orientation between the magnetic field and the density \ita{gradient}.}.\\
Fig.~\ref{fig:terms_ind} shows the density dependence of each individual, averaged coefficient of 
Eq.~\eqref{eq:hro}. At low densities, the $A$--coefficients are both 
negative. For the weakly magnetised cases with compressive driving 
they stay almost constant as a function of density. Generally, the 
$A_{23}$--term is larger in magnitude at the highest densities, 
indicating that here the shear in the flow dominates. Note that 
the $A_1$--term transitions to positive values for \ita{B1com}, but does 
not become dominant. The corresponding solenoidal cases show no 
transition to positive values, but instead a further decrease of the 
$A_{23}$--term. \\
As stated above, only the runs with initial $\beta=0.01$ transition 
towards a preferentially parallel alignment. This is due to the 
fact that $A>C$ and that $A$ becomes positive at some density. The 
earlier change in relative orientation for the compressive case 
matches well with the earlier transition of the $A$--coefficient. 
To be more precise, it is the transition of the $A_1$--term and 
the subsequent dominance of it. The respective $A_{23}$--term 
becomes positive at almost an order of magnitude higher densities. 
\bk{The behaviour of the individual $A$--terms implies that the gas 
undergoes strong shocks and that these shocks are more dominant 
than shear in the fluid flow.}
In contrast, this large difference in density is not observed for 
the solenoidal scenario, because here the overall magnitude of the 
velocity divergence is small.
In agreement with \citet{seifried2020}, the coefficients responsible for 
the change in relative orientation might depend on the specific 
physical conditions. This is further illustrated by the 
evolution of $\zeta_\mathrm{HRO}$ for the run \ita{B0.01com}. $\zeta_\mathrm{HRO}$ reaches a value of zero, but does not 
proceed to negative values. Instead, it increases again, before it 
finally decreases to negative values. As can be seen in 
Fig.~\ref{fig:terms_ind}, the coefficient $C$ takes non-zero values 
and does become non-negligible at $n>10^4\,\mathrm{cm}^{-3}$. It 
thus starts to impact the evolution up to the point where 
$A_1$ becomes completely dominant.\\
In summary, the presented figures reveal that the behaviour of the 
flow field in a 
strongly magnetised medium is vital for the details of the change 
in relative orientation. 
\subsection{Time evolution of the HRO parameter}
\begin{figure*}
    \centering
    \begin{tabular}{ll}
    \includegraphics[width=0.36\textwidth,angle=-90]{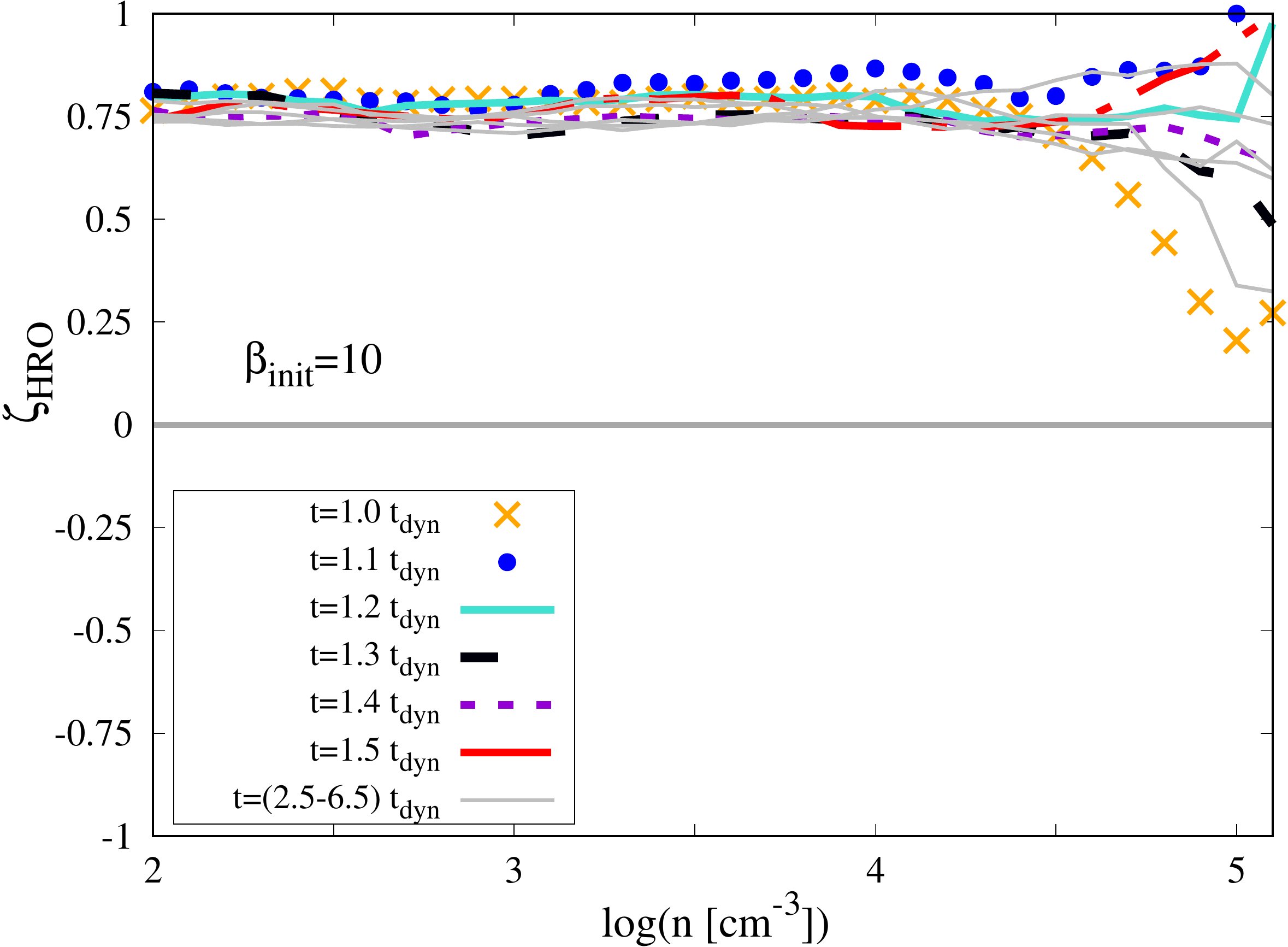}&\includegraphics[width=0.36\textwidth,angle=-90]{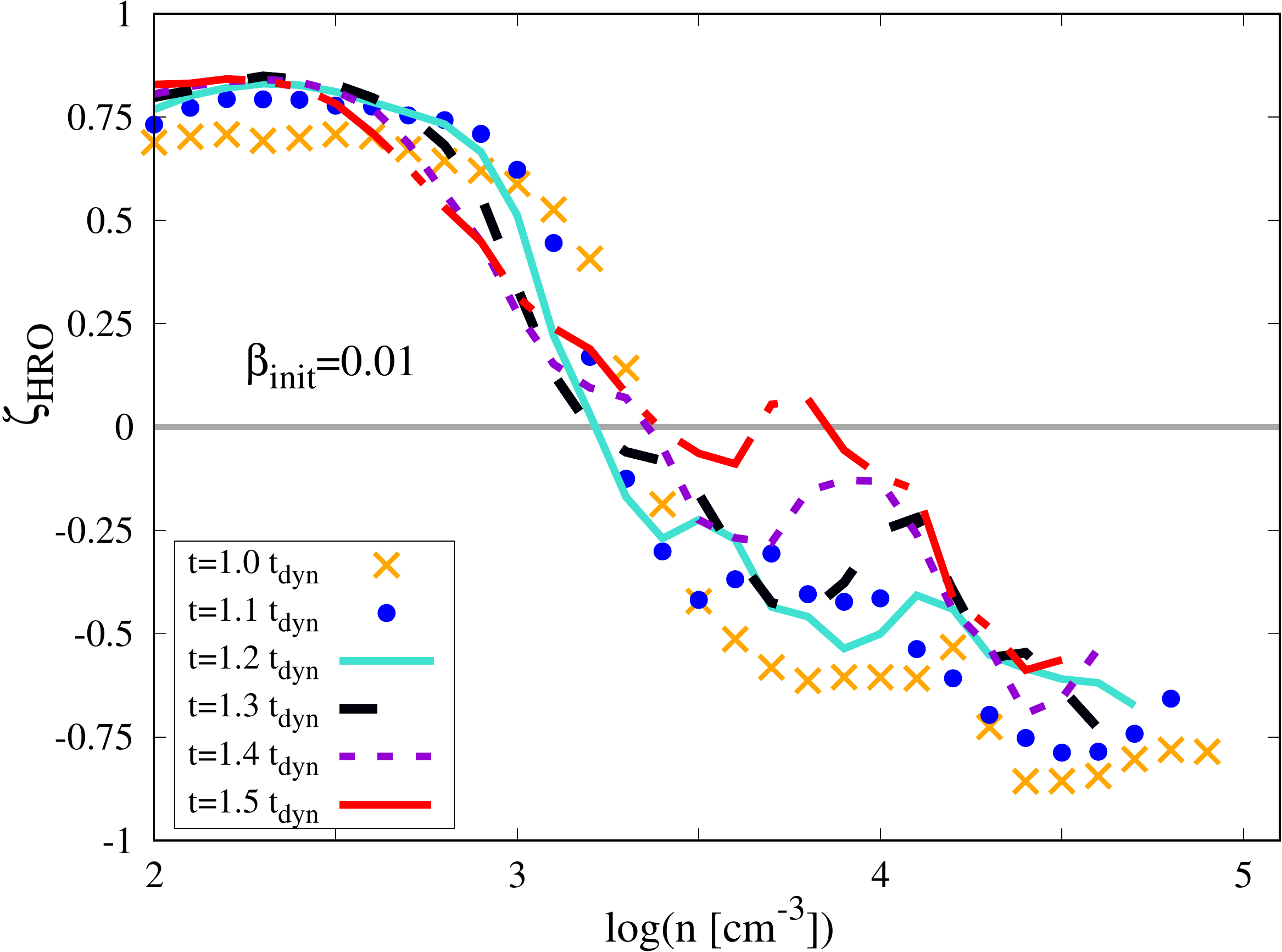}
    \end{tabular}
    \caption{Relative orientation parameter, $\zeta_\mathrm{HRO}$, for 
    simulation \ita{B10com} (left) and \ita{B0.01com} (right) as a function of density for various times between $t=t_d$ and $t=1.5\,t_d$. In case of run \ita{B10com}, 
    we also show very late stages up to $t=6.5\,t_d$ (thin grey lines). Apart from some expected temporal fluctuations, the relative orientation between the 
    magnetic field and the density gradient reaches a quasi-stationary 
    state.
    }
    \label{fig:time_hro}
\end{figure*}

Here, we investigate the time evolution of the relative orientation parameter. 
For this, we show in Fig.~\ref{fig:time_hro} the shape parameter for times between one and 1.5 dynamical times. Usually, the turbulence is expected to have reached a quasi-stationary 
state around one dynamical time. As is clearly seen, both the low and high magnetisation 
cases reach a stationary configuration, where only small fluctuations are seen. These 
latter are expected as the system stays supersonically turbulent. In general, both 
systems show that the relative orientation depends on the magnetisation of the gas and 
does not change significantly over time, once the final state is reached. This is 
even more important for the low magnetisation case, as it shows that this system will 
never be able to transition towards a preferentially parallel alignment of the 
magnetic field and the density gradient.\\
We furthermore show the time evolution of the governing parameters in Eq.~\eqref{eq:hro} in 
Figs.~\ref{fig:time_covera} and \ref{fig:time_terms}. As expected, the governing 
terms do not change much with time. All fluctuations are short-periodic around a 
certain mean value.
\subsection{Connection to observations}
\begin{figure}
    \centering
    \includegraphics[width=0.25\textwidth,angle=-90]{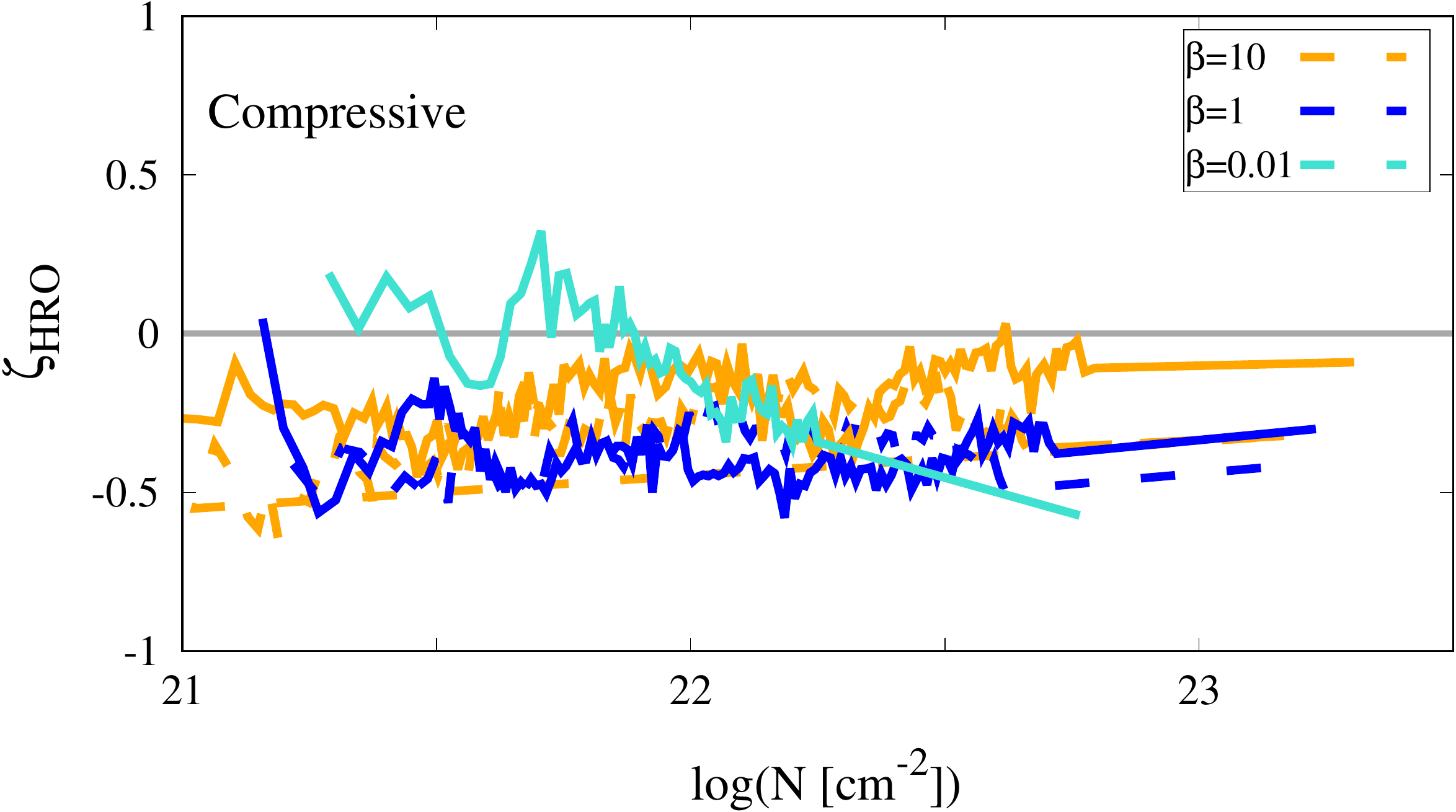}\\
    \includegraphics[width=0.25\textwidth,angle=-90]{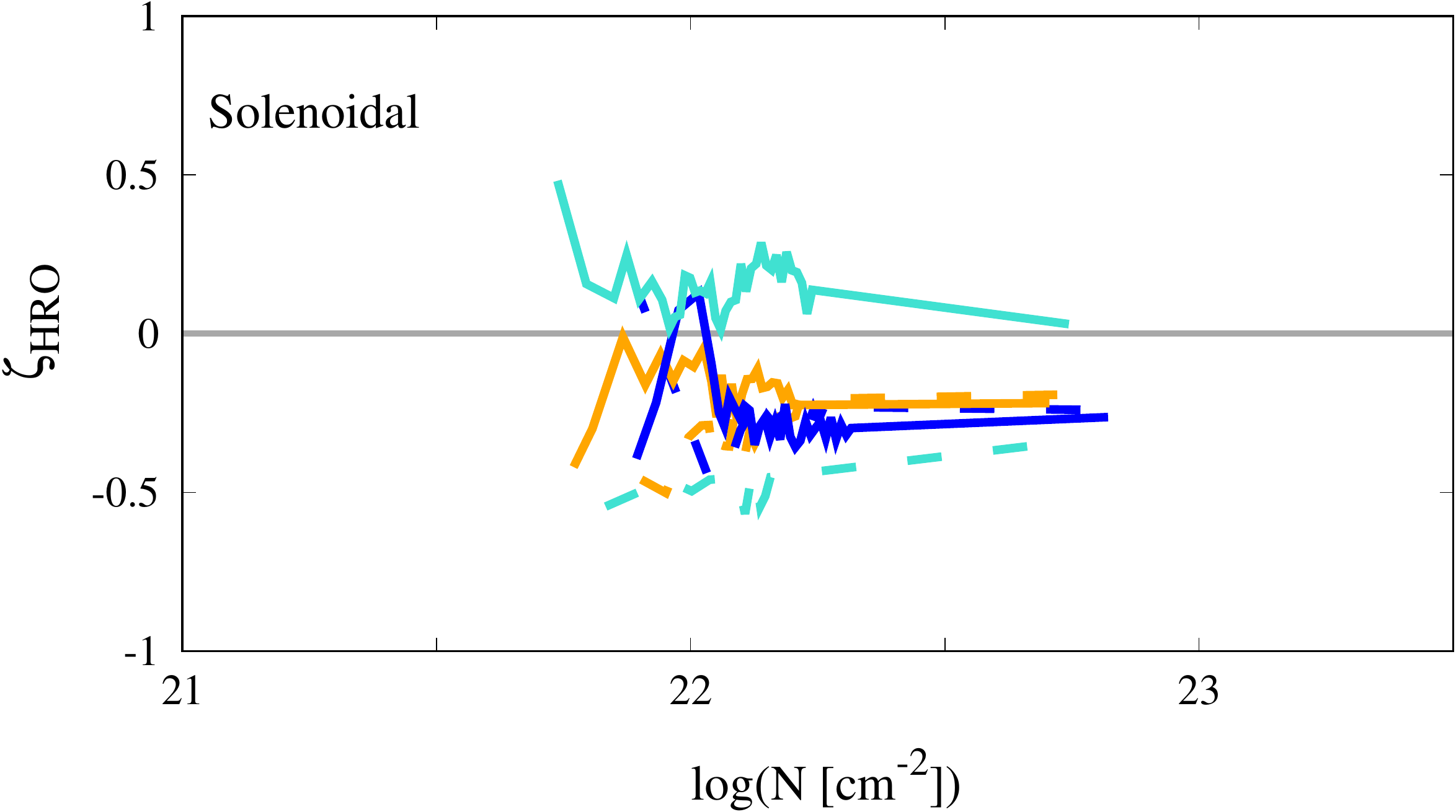}\\
    \includegraphics[width=0.25\textwidth,angle=-90]{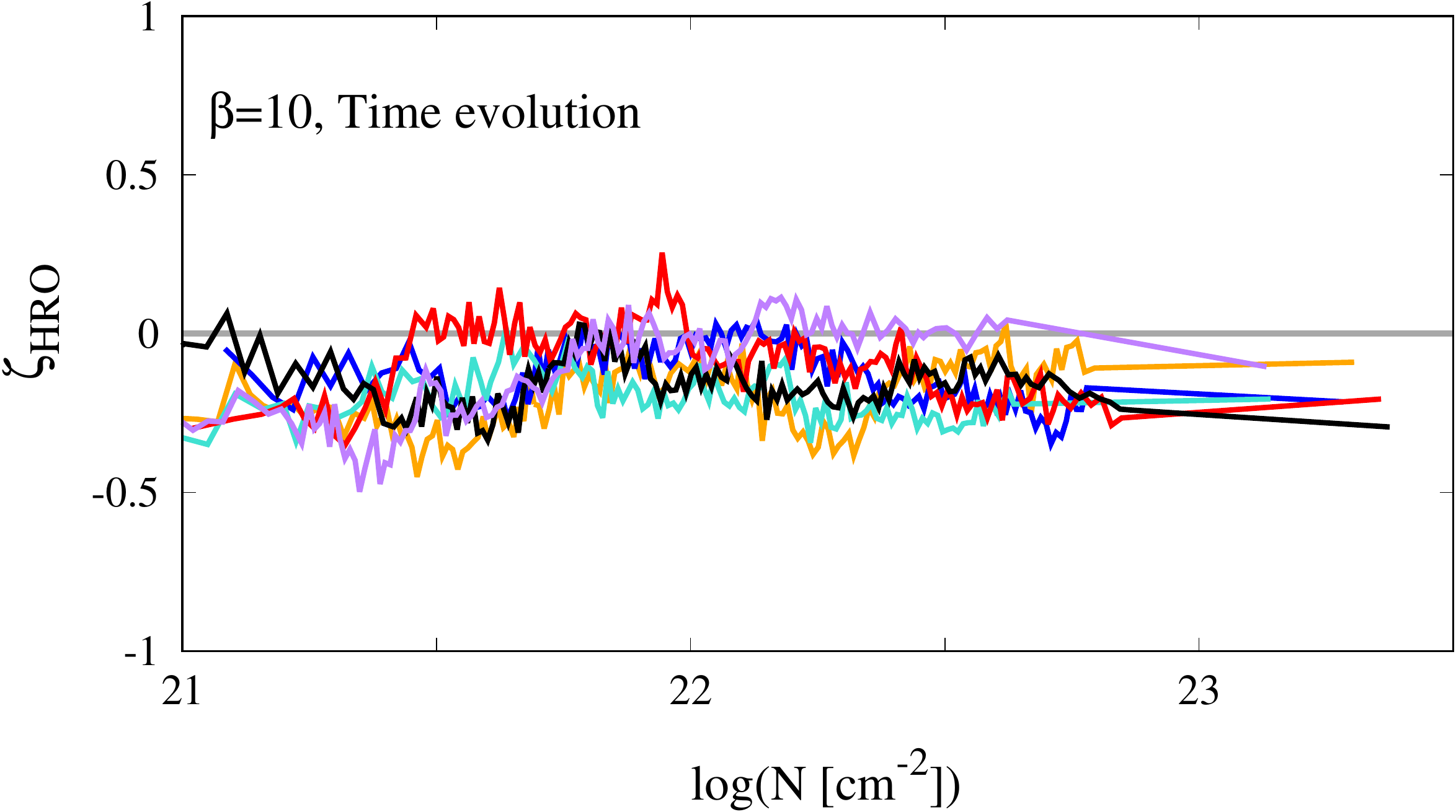}
    \caption{Shape parameter as a function of logarithmic 
    column density for projections along the initial 
    background magnetic field direction (dashed) and 
    perpendicular to it (solid) for both compressive (top) 
    and solenoidal (middle) turbulence driving. Colour denotes different initial magnetisation of the gas. Note that 
    there are no data shown for integration along the 
    x-direction for run \ita{B0.01com} due to the too small 
    column density range. The bottom panel shows the 
    time evolution of the shape parameter for run 
    \ita{B10com}. Although the shape parameter never 
    attains positive values over the entire 
    column density range, it is evident that it 
    fluctuates between almost no preferred 
     and a slightly parallel orientation.}
    \label{fig:hro_cdens}
\end{figure}

Lastly, we bridge the gap between our theoretical 
calculations and observations. For this task, we calculate  
the pseudo Stokes \bk{components} \citep[here given only for integration 
along the x-direction,][]{Fiege00}
\beq
q=\int{n\frac{B_\mathrm{z}^2-B_\mathrm{y}^2}{B^2}\mathrm{d}x},
\eeq
\beq
u=2\int{n\frac{B_\mathrm{z}B_\mathrm{y}}{B^2}\mathrm{d}x}.
\eeq
From these we calculate the polarisation angle 
\beq
\chi=\frac{1}{2}\mathrm{arctan}\left(\frac{u}{q}\right)
\eeq
and finally generate a pseudo polarisation vector 
\mbox{$\vek{p}=p_0\left(\mathrm{sin}\chi\, \hat{\vek{e}}_x+\mathrm{cos}\chi\,\hat{\vek{e}}_y\right)$}, where 
the $\hat{\vek{e}}_i$ vectors are the unit vectors along the directions perpendicular to the line of sight and $p_0=0.1$ 
is the maximum polarisation fraction.\\
The resulting HRO shape parameter as a function of gas 
column density is shown in Fig.~\ref{fig:hro_cdens}. Comparison of these data 
with the shape parameter as a function of volume density in Fig.~\ref{fig:hro} 
reveals the large impact of projection effects along the line of sight for the 
scenarios with low initial magnetisation. Here, the orientation is seen to be 
parallel for the entire column density range. We remind the reader that in 3D it 
is actually vice versa, i.e. entirely perpendicular alignment between the 
magnetic field and the density gradients. We further find only 
minor differences between the runs with an initial 
magnetisation of $\beta=10$ and $\beta=1$ or between the 
lines of sight. A major difference appears when the gas is 
magnetically dominated. Here, the data reveal no preferred 
orientation at the lower column densities and an increasing 
alignment of the two fields at higher column densities. 
Interestingly, this decreasing trend is not observed as clearly
in the solenoidal counterpart. However, here a clear 
difference is evident for the two different lines of sight. 
Whereas the integration perpendicular to the initial field 
reveals no preferred or at most a slight perpendicular 
alignment, the corresponding projection along the initial 
background field shows a pronounced parallel alignment. 
In addition, similar to the compressive case, no difference is 
seen for the low magnetisation cases. \\
The negative relative orientation parameter has also 
been found recently in synthetic observations by \citet{seifried2020} for weakly magnetised media. We show in the 
bottom panel of Fig.~\ref{fig:hro_cdens} the retrieved shape 
parameter for the times already provided in Fig.~\ref{fig:time_hro}. Whereas there appeared almost no 
fluctuations across a large density range in the volume density 
calculation, the corresponding projections strongly fluctuate 
both as a function of column density and time. In fact, the most 
striking feature is the absence of a preferred perpendicular 
alignment in projection. Instead, $\zeta_\mathrm{HRO}\sim0$ 
seems to be the largest typical value in projection. This 
indicates that there appears no preferred orientation. 
Interestingly, the red line is observed to be quite similar 
to the case of a compressive turbulent velocity field and a 
strong magnetic field, again emphasising the role of 
projection effects. It is hence hard, if not 
impossible, to differentiate between compressive or solenoidal 
turbulence driving via the relative orientation of the 
magnetic field and (column-) density gradient.

\section{Summary and conclusions}\label{sec:summary}

We presented a study on the relative orientation between the magnetic field and the gas density gradient in non-gravitating, turbulent media with varying initial 
magnetisation. 
We considered numerical simulations where the turbulence is driven by an external forcing term either in a fully compressive or entirely solenoidal way. 
Our case study thus covers two extreme conditions of supersonic interstellar turbulence.

Our most important finding is that compressive o solenoidal turbulence alone does not induce a change in relative orientation between the magnetic field and the gas density gradient.
The change in relative orientation reported in previous numerical studies that include self-gravity is only found in the simulations with the highest magnetization in our data set.
The nature of the turbulent forcing of the velocity field only mildly affects the transition in the relative orientation.
This suggests that, in agreement with previous numerical studies, the change in relative orientation primarily depends on the magnetisation of the gas.

The configuration where the magnetic field parallel to the gas density gradient, or parallel to the isodensity contours, is only observed clearly in the simulation with high magnetization and compressive turbulence.
This is in agreement with the interpretation presented in \cite{soler&hennebelle2017}, where the main driver of a change in relative orientation between the magnetic field and the density structures is the compression of the gas.

We studied the time evolution of the simulations and found that the relative orientation between the magnetic field and the gas density gradient does not significantly change within one dynamical time.
In this quasi-steady state, the relative orientation depends only on the initial magnetisation of the gas.
We conclude that a \ita{true} change in relative orientation can only be 
achieved in a medium with a dynamically significant magnetic 
field.


\section*{Acknowledgement}

The authors acknowledge Paris-Saclay University's Institut Pascal program ``The Self-Organized Star Formation Process'' and the 
Interstellar Institute for hosting discussions that nourished the development of the ideas behind this work. BK enjoyed 
discussions with L.~Fissel, S.~E.~Clark and C.~Federrath.
JDS thanks R.~Pudritz and E.~Ostriker for the conversations that encouraged to this work.

BK thanks for funding from the DFG grant BA~3706/15-1 and via the Australia-Germany Joint Research Cooperation Scheme (UA-DAAD). 
JDS acknowledges funding from the European Research Council under the Horizon 2020 Framework Program via the Consolidator Grant 
CSF-648505.
The simulations were run on HLRN-III under project grant hhp00043. 
The \textsc{flash} code was in part developed by the DOE-supported ASC/Alliance Center for Astrophysical Thermonuclear Flashes at 
the University of Chicago.\\

\section*{Data availability}
The data underlying this article will be shared on reasonable request to the corresponding authors.

\begin{appendix}
\section{Time evolution of coefficients}
\begin{figure}
    \centering
    \includegraphics[width=0.35\textwidth,angle=-90]{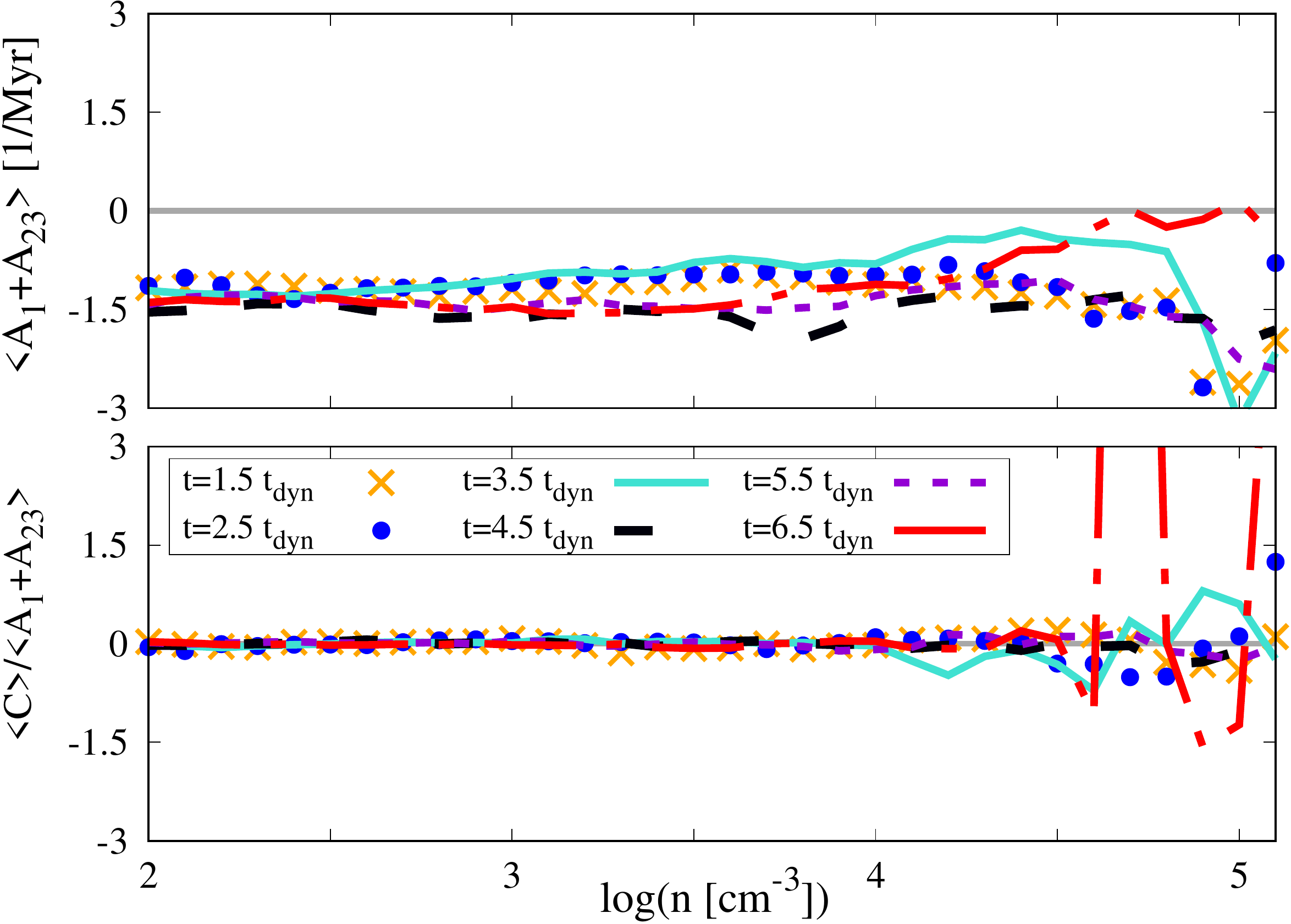}
    \caption{Same as Fig.~\ref{fig:terms_tot_ratio}, but for the time evolution of only run \ita{B10com}. Once, the system has 
    reached a quasi-steady state, the parameters, 
    which govern the rate of change of the angle 
    between the magnetic and density gradient fields,  
    do not vary much.}
    \label{fig:time_covera}
\end{figure}

\begin{figure}
    \centering
    \includegraphics[width=0.35\textwidth,angle=-90]{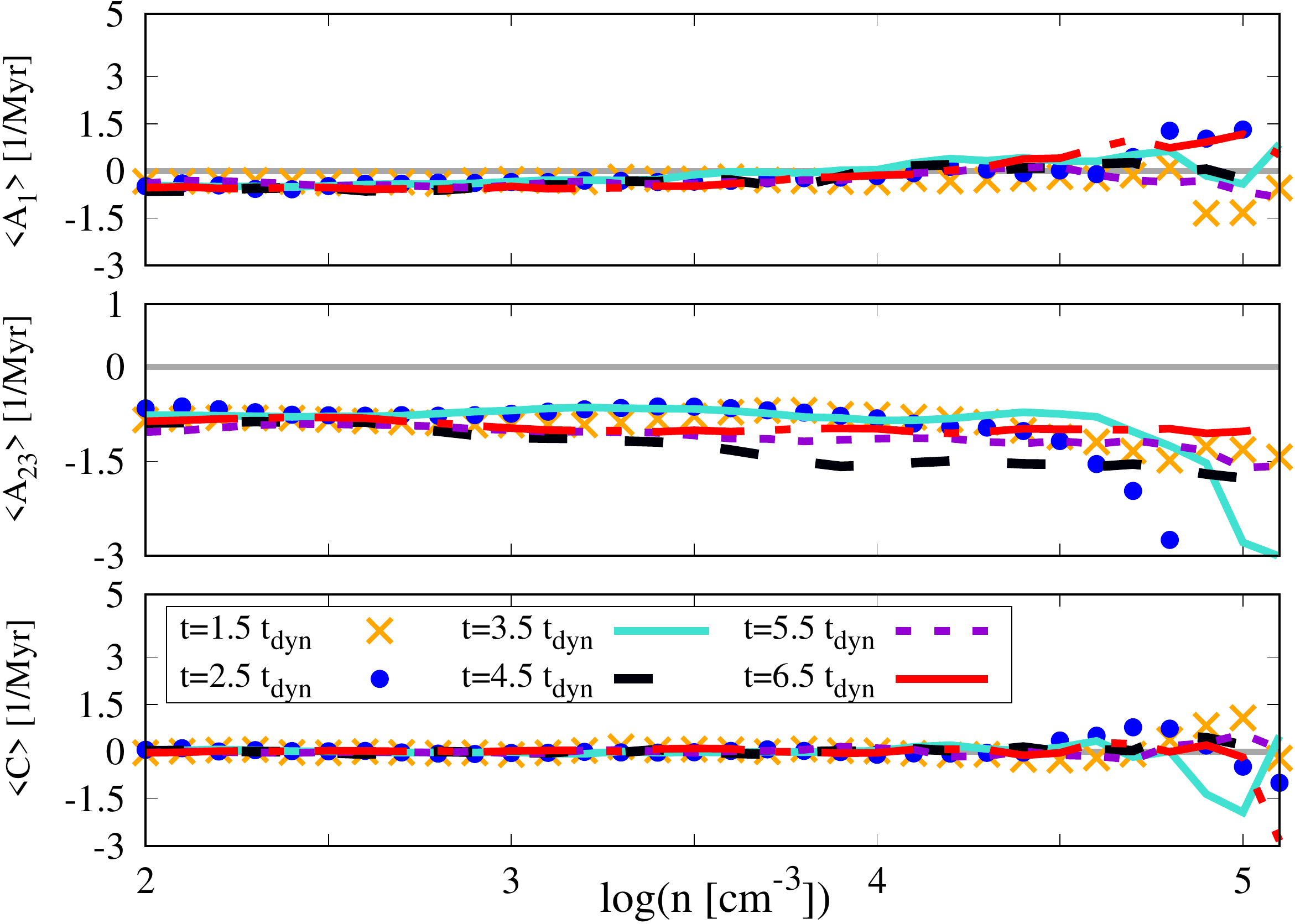}
    \caption{Density dependence of the individual 
    parameters, which determine the rate of change of 
    the relative orientation for run \ita{B10com}. Different lines indicate 
    different times. Once a quasi-stationary state is 
    reached, the relative orientation does not change 
    anymore.}
    \label{fig:time_terms}
\end{figure}

As a support to our analysis, we provide in Figs.~\ref{fig:time_covera} and \ref{fig:time_terms} the time evolution of the governing coefficients in eq.~\ref{eq:hro}. These data show that the solutions are time independent and 
that such weakly magnetised systems will never manage to change its relative 
orientation.
\end{appendix}

\bibliography{astro}

\bibliographystyle{mn2e}

\end{document}